\documentclass[aip, amsmath, amssymb, preprint]{revtex4-2}

\usepackage{graphicx}
\usepackage{dcolumn}
\usepackage{bm}
\usepackage{ifpdf}

\usepackage[T1]{fontenc}
\usepackage{mathptmx}

\begin{document}

\title{Deterministic positioning of nanophotonic waveguides around single self-assembled quantum dots}

\author{T. Pregnolato}
\altaffiliation[Currently at: ]{Department of Integrated Quantum Technology, Ferdinand-Braun-Institut, Leibniz-Institut f\"{u}r H\"{o}chstfrequenztechnik, Gustav-Kirchhoff-Stra{\ss}e 4, 12489 Berlin, Germany}
\affiliation{Center for Hybrid Quantum Networks (Hy-Q), Niels Bohr Institute, University of Copenhagen, Blegdamsvej 17, DK-2100 Copenhagen, Denmark}%

\author{X.-L. Chu}%
\affiliation{Center for Hybrid Quantum Networks (Hy-Q), Niels Bohr Institute, University of Copenhagen, Blegdamsvej 17, DK-2100 Copenhagen, Denmark}%

\author{T. Schr\"{o}der}%
\altaffiliation[Currently at: ]{Department of Physics, Humboldt-Universit\"{a}t zu Berlin, Newtonstr. 15, 12489 Berlin, Germany}
\affiliation{Center for Hybrid Quantum Networks (Hy-Q), Niels Bohr Institute, University of Copenhagen, Blegdamsvej 17, DK-2100 Copenhagen, Denmark}

\author{R. Schott}%
\affiliation{Lehrstuhl f\"{u}r Angewandte Festk\"{o}rperphysik, Ruhr-Universit\"{a}t Bochum, Universit\"{a}tsstrasse 150, D-44780 Bochum, Germany}

\author{A. D. Wieck}%
\affiliation{Lehrstuhl f\"{u}r Angewandte Festk\"{o}rperphysik, Ruhr-Universit\"{a}t Bochum, Universit\"{a}tsstrasse 150, D-44780 Bochum, Germany}

\author{A. Ludwig}%
\affiliation{Lehrstuhl f\"{u}r Angewandte Festk\"{o}rperphysik, Ruhr-Universit\"{a}t Bochum, Universit\"{a}tsstrasse 150, D-44780 Bochum, Germany}

\author{P. Lodahl}%
\email{lodahl@nbi.ku.dk}
\affiliation{Center for Hybrid Quantum Networks (Hy-Q), Niels Bohr Institute, University of Copenhagen, Blegdamsvej 17, DK-2100 Copenhagen, Denmark}%

\author{N. Rotenberg}
\email{nir.rotenberg@queensu.ca}
\altaffiliation[Currently at: ]{Department of Physics, Engineering Physics and Astronomy, Queen's University, Kingston, K7L 3N6, Canada}
\affiliation{Center for Hybrid Quantum Networks (Hy-Q), Niels Bohr Institute, University of Copenhagen, Blegdamsvej 17, DK-2100 Copenhagen, Denmark}%

\date{\today}

\begin{abstract}
The capability to embed self-assembled quantum dots (QDs) at predefined positions in nanophotonic structures is key to the development of complex quantum photonic architectures.  Here, we demonstrate that QDs can be deterministically positioned in nanophotonic waveguides by pre-locating QDs relative to a global reference frame using micro-photoluminescence ($\mu$PL) spectroscopy. After nanofabrication, $\mu$PL images reveal misalignments between the central axis of the waveguide and the embedded QD of only $(9\pm46$)~nm and $(1\pm33$)~nm, for QDs embedded in undoped and doped membranes, respectively. A priori knowledge of the QD positions allows us to study the spectral changes introduced by nanofabrication. We record average spectral shifts ranging from 0.1 to 1.1~nm, indicating that the fabrication-induced shifts can generally be compensated by electrical or thermal tuning of the QDs. Finally, we quantify the effects of the nanofabrication on the polarizability, the permanent dipole moment and the emission frequency at vanishing electric field of different QD charge states, finding that these changes are constant down to QD-surface separations of only 70 nm. Consequently, our approach deterministically integrates QDs into nanophotonic waveguides whose light-fields contain nanoscale structure and whose group index varies at the nanometer level.
\end{abstract}

\maketitle

\section{Introduction}
The rapid maturation of the InAs self-assembled quantum dot (QD) platform, and in particular the ability to interface these emitters with high quality nanophotonic structures,~\cite{Lodahl2015} has opened up viable routes towards the creation of integrated single-photon sources~\cite{Shields2007} for quantum network applications.~\cite{Borregaard2018a, Lodahl2018}  This increasing viability of QD-based photonic technology can be traced to three milestones in the field: the growth of high quality QDs via the Stranski-Krastanov technique,~\cite{Petroff2001} the ability to couple emission from QDs to photonic modes with near-unity efficiency,~\cite{Arcari2014, Ding2016} and the use of doped heterostructures to charge stabilize the environment of the emitters.~\cite{Kuhlmann2015, Lobl2017}  Altogether, these allow for efficient and highly coherent light-matter interactions~\cite{Lobl2017, Thyrrestrup2018, Turschmann2019} and the generation of highly-indistinguishable photons,~\cite{Kirsanske2017, He2013} which are basic capabilities of quantum-photonic processing elements.

To date, the vast majority of QD-based devices are fabricated with no \emph{a priori} spatial or spectral knowledge about the emitters. This lack of information results in low yields when QDs are interfaced with nanoscale or dispersive elements such as waveguides or resonators, precluding the scaling up of these systems into complex architectures.  Furthermore, without prior knowledge of the QD properties, it has not been possible to quantify the effects of nanofabrication techniques on individual emitters.

In order to address these issues, a variety of techniques have recently been developed. Although differing in the specific strategies, they all present two main steps: first, the QDs are located in bulk samples; then photonic structures are deterministically fabricated about the detected positions. Interestingly, most of the reported works have primarily been concerned with improving the precision $\delta$ with which the emitters are located, whereas it is the final QD-nanostructure misalignment $\Delta$ that affects the performances of the fabricated devices, especially in those where the electromagnetic field is strongly confined and varies spatially. For photonic crystal waveguides, for example, the relevant length scale is the Bragg wavelength inside the medium. This corresponds to an effective QD emission wavelength in the medium of $\lambda/2n\approx130~\text{nm}$ for a typical QD emission wavelength of $910\text{~nm}$ and refractive index $n\approx3.5$. In more quantitative terms, an alignment precision of $50\text{~nm}$ in a photonic-crystal waveguide implies that a coupling efficiency~$\beta\ge~96\%$ can be achieved deterministically.~\cite{Javadi2017} 

One approach is to use \emph{in-situ} techniques, where cathodoluminescence (CL)~\cite{Gschrey2013, Gschrey2015, Kaganskiy2015} or micro-photoluminescence ($\mu$PL)~\cite{Dousse2009, Sartison2017} spectroscopy first locates the QDs followed by electron-beam or photolithography to pattern the photonic elements.  With CL, QDs were positioned within a nanoscale multimode beam-splitter, with a QD-nanostructure misalignment of only $\Delta$~=~34 nm,~\cite{Schnauber2018} while photolithography defines structures with micron dimensions and therefore does not require the same degree of accuracy.

An alternative approach is to first locate the QDs relative to alignment markers, using either scanning electron microscopy~\cite{Badolato2005} or $\mu$PL,~\cite{Thon2009, Kojima2013, Sapienza2015, Coles2016} then fabricate structures in a separate step.  Separating the localization and nanofabrication has the distinct advantage of parallelizing the deterministic nanofabrication procedure and is therefore more compatible with the design of time-intensive, complex lithography masks that contain many elements. The images taken in these protocols contain both the \emph{emission} from the QDs and the \emph{reflection} from the alignment markers, yielding typical localization precision of $\delta<$ 10 nm.~\cite{Liu2016, He2017}  The reflected image, however, depends on the excitation angle, while the emission pattern from the QDs does not.  Hence, slight misalignment of the excitation beam introduces alignment errors, which are reflected in the much-larger QD-nanostructure misalignments $\Delta$ found. For example, in the case of photonic crystal cavities in strong coupling regime,~\cite{Thon2009,Kojima2013} final QD-nanostructure misalignment of the order of $\Delta\approx50$~nm has been determined by comparing the measured coupling constant with the maximum calculated value. In another example, misalignments $\Delta$ between 50 and 250 nm in circular Bragg gratings have, instead, been inferred from simulations.~\cite{Sapienza2015} Finally, for other devices, such as micropillars, that have been deterministically integrated with QDs, no value of $\Delta$ has been reported; in these cases, the device diameter - typically up to a few $\mu$m's large~\cite{Liu2016a, He2017} - places an upper bound on the misalignment $\Delta$. Furthermore, the lack of a systematic study of these misalignments means that it is not possible to determine whether the errors arise due to optical measurements or the subsequent nanofabrication.

In contrast, we present the first quantitative and statistical study of the alignment of pre-located QDs to nanoscale photonic waveguides, using a straightforward improvement to the existing $\mu$PL protocols. Our method results in a systematic misalignment of 9 and 1 nm, randomly distributed with standard deviations of 46 and 33 nm, for samples where the QDs cannot and can be electrically controlled, respectively. We furthermore study the effect of fabricating quasi-one-dimensional nanophotonic waveguides or two-dimensional photonic-crystal waveguides (PhCWs) on the spectral response of the QDs, and on the different exciton complexes.

\section{Prelocalization of quantum dots}
In this work we set out to achieve a final QD-nanostructures misalignment $\Delta< 50$~nm. To do so we take the following steps: (i) Fabrication of a grid of alignment markers on top of a wafer containing QDs; (ii) Localization of QDs within each grid square and measurement of the spectral properties; (iii) Fabrication of nanostructures at preselected positions. To localize the QDs, in step (ii), we modify the protocol of \textcite{Liu2016} in a manner that yields the desired final accuracy.  We use only photoluminescence, and not reflection, to image both the alignment markers and the QDs.

\begin{figure}
\begin{center}
\includegraphics[width=8.4cm]{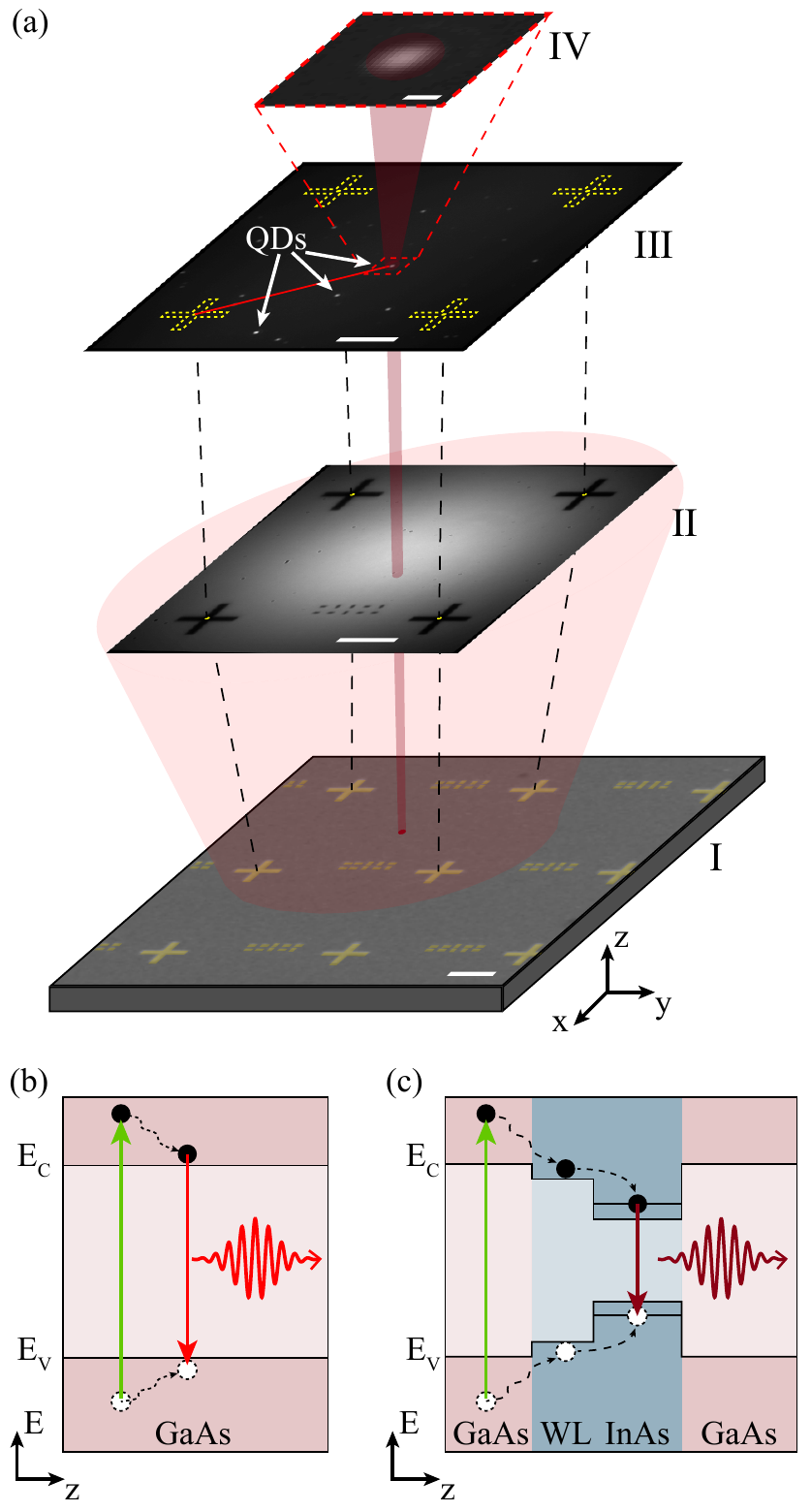}
\caption{Prelocalization protocol for epitaxially grown QDs (real data shown). (a) I: False color SEM of the GaAs membrane with embedded InAs QDs, with a gold alignment mask of crosses. II: First the alignment mask is imaged and the crosses located, using the photoluminescence of the membrane, as discussed in the main text. III: The QD photoluminescence is imaged and spatially correlated with the cross positions. As example, we point to several QDs. IV: Closed-up image of the emission pattern of a single QD.  All scalebars are 10~$\mu$m in length, except for IV, where it is 500~nm long.  Energy levels of the GaAs membrane (b) and embedded QD and wetting layer WL~(c).  In both cases, above band illumination (green arrow) and subsequent fast, non-radiative decay (black arrows) excites an electron (black symbol) to the conduction band, leaving a hole (white symbol) in the valance band of the GaAs or QD. Recombination of the electron and hole results in the emission of a photon, whose wavelength is typically near 830 - 850~nm and 930~nm for the membrane and QDs (light and dark red arrows, respectively).}
\label{fig:f1_concept}
\end{center}
\end{figure}

Our global coordinate frame is set by a grid of gold crosses, which we fabricate on top of a GaAs membrane with embedded InAs QDs, as shown in Fig.~\ref{fig:f1_concept}a.  In this study, we use two types of wafers: the QDs are embedded either within an undoped GaAs membrane that is 160-nm thick, or within a p-i-n diode.~\cite{Kirsanske2017} In either case, each square within the grid is 40~$\mu$m by 40~$\mu$m and is identified by a binary label (set of small gold rectangles fabricated with different orientations).  We also fabricate a grid of solid gold lines that we use to quantify and correct for the rotation of our images (not shown).

We locate the position of the QDs within each square of the grid in two steps, first imaging the crosses and then finding the QDs, all at a cryogenic temperature of about 10 K. Examples of these two images are shown in panels II and III of Fig.~\ref{fig:f1_concept}a. Although the excitation scheme does not change -- namely we use above band excitation at 780 nm, illuminating in widefield configuration an area slightly larger then the square -- our imaging protocol depends on the type of wafer used.  In either case we image the reference markers and the QDs separately, using emission from the sample in both cases, and not via the reflected laser light.  Consequently, no alignment errors are introduced due to slight angles of the excitation beam.  This improvement simplifies the measurement, and, as we show below, decreases the final misalignment between the QDs and nanostructures below previously reported values.

For the intrinsic sample, we use an 800 nm long-pass filter to block the reflected laser light, using the luminescence of the GaAs membrane to image our reference markers (using an Andor iKon-M CCD camera). As shown in Fig.~\ref{fig:f1_concept}a, panel II, the gold markers block the light emitted by the GaAs and therefore appear as shadows in the resultant image.  By fitting a line to each arm of the crosses, we find their centers, typically with an accuracy of 3.1 nm that is largely determined by the signal-to-noise ratio of the measurement. This value is far below the diffraction limit of the setup, and is typical of what is reported in the literature.~\cite{Liu2016}

Taking separate images of the alignment markers and QDs has an additional advantage. To locate the QDs, a narrowband filter $\left(935\text{~nm}\pm0.5\text{~nm}\right)$ is placed in the collection path, blocking the emission from the GaAs, resulting in background-free images that contain 10-20 well-separated QDs, as shown in panel III of Fig.~\ref{fig:f1_concept}a. We then fit a two-dimensional Gaussian to each QD, finding its location to within 0.6 nm that, due to the higher signal to background ratio, is three times smaller than the value reported in current state-of-the-art literature.~\cite{Liu2016} Assuming that our field-of-view is unchanged between these two images, we find the position of each QD relative to the global reference frame with an accuracy $\delta~=~4.9$~nm that is dominated by the uncertainty in the cross position.

The ability to electrically control the optical properties of the QDs on the gated sample allows us to further improve the localization protocol, a change that ultimately leads to better QD-structure alignment accuracy (see Sec.~\ref{sec:integration}).  Here, the QDs are embedded in a diode with a large built-in field, meaning that the energy levels are strongly tilted relative to those of Fig.~\ref{fig:f1_concept}c.  Consequently, in this natural state (i.e. with no applied bias voltage) the excited electrons quickly tunnel out of the QD, and no emission is observed.  Conversely, we can `turn on' the QDs by applying a bias voltage to recover the configuration of Fig.~\ref{fig:f1_concept}c.~\cite{Fry2000a} This electrical control allows imaging both the crosses and the QDs independently, without changing filters but rather by tuning the applied voltage, ensuring that our field-of-view is constant.  To do so, a 900 nm long-pass filter is placed in the collection path, blocking the relatively strong emission from the GaAs, which would otherwise swamp the signal from the QDs.  Instead, we use the tail end of the emission from the wetting layer to image the crosses with the QDs `off', when we do not apply a bias voltage. In these images we find the position of the center of each cross with an accuracy of 5.5 nm.  We then apply 500~mV to turn on the QD emission, which then dominates over the wetting layer fluorescence. The resultant image can be used to localize the emitters' positions to within 0.7 nm.  As before, by correlating the two images we find the absolute position of each QD with an accuracy $\delta~=~9.2$~nm. This value is larger than the one measured for the intrinsic sample due to a larger signal-to-background ratio, which is caused by the dimmer signal collected from the wetting layer. Nonetheless, we note that this value, which is consistent with earlier reports,~\cite{Liu2016, He2017} is below the level of overlay accuracy ($\pm15~\text{nm}$) that our e-beam lithography system (Elionix ELS-F125) can achieve.

After localization, and if desired, additional measurements can be made using pre-selected QDs.  Here, for example, we switch to a confocal setup, where we can excite each QD individually and record its emission spectrum (see Sec.~\ref{sec:Effects} below).  Similarly, the lifetime or single-photon purity of each emitter can be quantified before fabrication.

\section{Deterministic integration of quantum dots into nanophotonic waveguides}\label{sec:integration}
\begin{figure}
\begin{center}
\includegraphics[width=8.4cm]{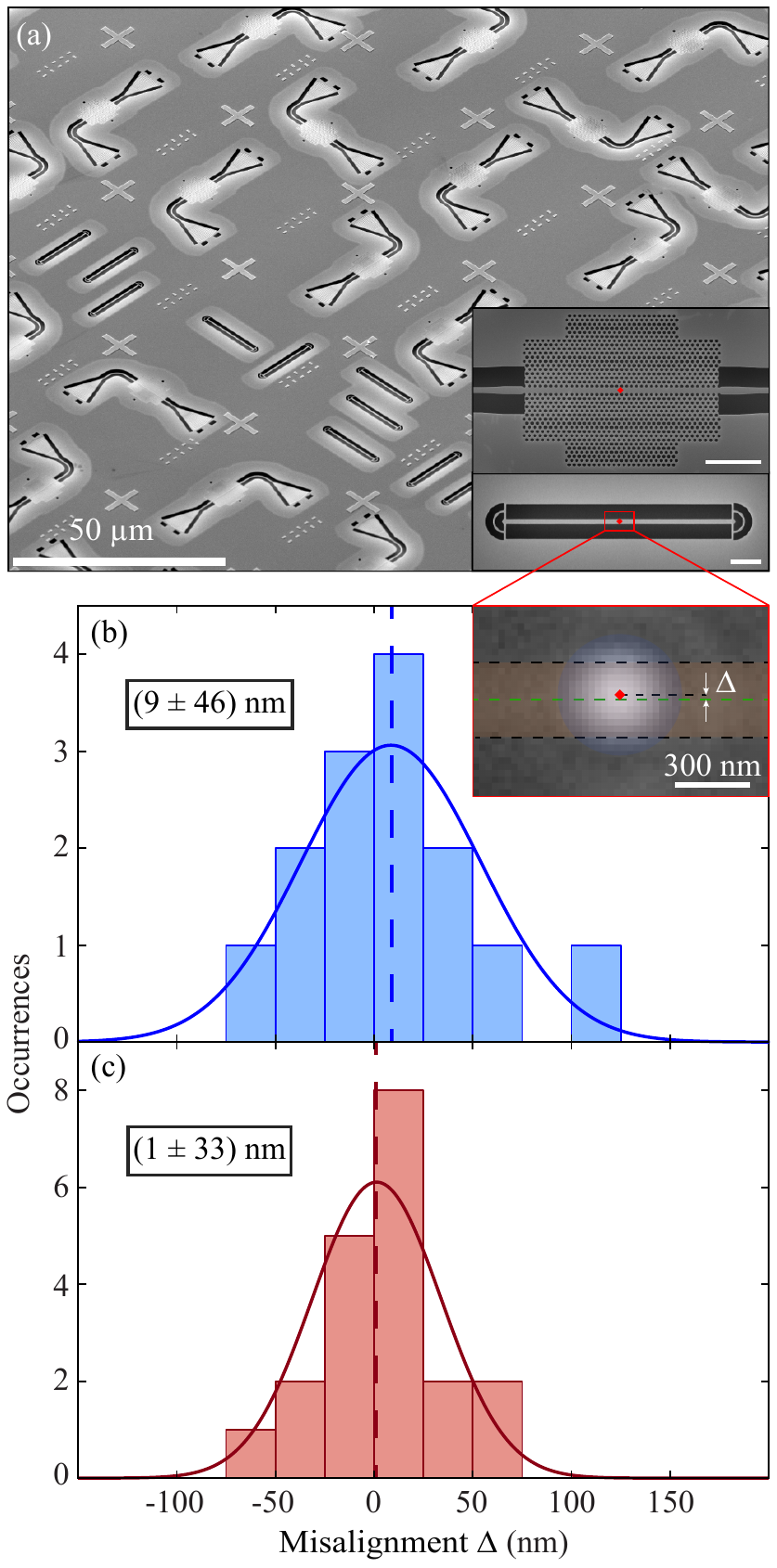}
\caption{Nanophotonic waveguides fabricated about pre-located QDs. (a) A representative area of the resultant photonic chips with both nanoguides and photonic crystal waveguides containing QDs (see inset, scalebars are both 2~$\mu$m). (b) Measured misalignment distribution between the QDs and the center of a nanoguide for the intrinsic sample.  The mean and standard deviation are given. The inset shows a false-color $\mu$PL image where the edges and center of the nanoguide are marked by black and green dashed lines, respectively, the QD emission is shown in blue with the center denoted by the red dot, and the relative misalignment is indicated by $\Delta$. (c) Same as in (b) but for the doped sample.  The smaller misalignment spread in (c) is attributed to the removal of the bandpass filter used in the localization of the QDs on the undoped sample, which ensures that both the QD and reference images are taken with the exact same field of view.}
\label{fig:f2_misalignment}
\end{center}
\end{figure}

The important figure-of-merit for the deterministic integration of solid-state emitters into nanophotonic elements is not the precision $\delta$ with which the emitters are located relative to a reference frame, but rather the final misalignment $\Delta$ between the emitters and the nanostructures.  We therefore fabricate suspended nanophotonic waveguides~\cite{Midolo2015} at the predetermined positions of selected QDs, as shown in Fig.~\ref{fig:f2_misalignment}a, making both quasi-one-dimensional nanoguides and two-dimensional PhCWs. The nanoguides have a rectangular cross-sections with widths ranging from ($288~\pm2$) to ($631\pm2$) nm, as measured from SEM images of the devices.  The PhCWs are created by removing a row of holes from photonic crystals with lattice constants ranging from 233 nm to 247 nm and hole radii ranging from 71 nm to 76 nm, and are fabricated with QDs at different positions within the photonic crystal unit cell (i.e. distances to the nearest surface), as we discuss below.

Micro-photoluminescence measurements on the waveguides after fabrication allow us to determine whether they contain the targeted QD and to subsequently quantify the misalignment $\Delta$ between the emitters and waveguides. The yield of the undoped sample, which contains 48 nanoguides and 50 PhCWs is $96\%$ and $92\%$ for the two types of structures, respectively. Here, the yield represents the fraction of waveguides that integrate the correct pre-selected QD, as determined by a cross-correlation study between the QD emission spectra collected before and after fabrication. In contrast, the yield of the doped sample is $93\%$ and $74\%$ for the nanoguides and PhCWs, respectively.  To understand the low yield of the doped PhCW structures we further break down the data by target distance of the QDs from the nearest surface, finding a yield of $94\%$ when this distance is $\geq 100$~nm but only $44\%$ for distances $< 100$~nm. In contrast, for the undoped sample, the yield was $80\%$ for distances $< 100$~nm, despite having a relatively higher misalignment, as we show below.  We can understand this difference as follows: The area near a lateral surface -- in our case the holes of the PhCW -- is known to have a reduced electrical conductivity due to a local depletion of the free carriers.~\cite{Berrier2007}  Hence we are unable to properly apply a bias voltage to QDs that are located in these regions, in contrast to QDs at similar separations from the hole surface in the undoped sample (which are always `on').  Regardless, we note that for both types of samples we succeeded in observing emission from QDs nominally positioned within about 30~nm of a hole edge.

The actual misalignment $\Delta$ between QD and nanostructure for each sample was quantified from images of the photoluminescence from the nanoguides, as shown in the inset to Fig.~\ref{fig:f2_misalignment}b.  In these images, the features of the nanostructures, whose width is below the resolution of our optical system, appear as three Gaussian peaks (see Fig.~\ref{fig:f9_fitWg} and Appendix~\ref{sec:Loc_Wgs} for further details). We estimate the center of the waveguide, marked with a green dashed line in the inset to Fig.~\ref{fig:f2_misalignment}b, from the fitted position of the central Gaussian.  Similarly, we find the position of the QD using a two-dimensional Airy function (center denoted by the red dot in Fig.~\ref{fig:f2_misalignment}), allowing to quantify the final misalignment $\Delta$ between the two.  A histogram of $\Delta$ for QDs embedded in the nanoguides on the intrinsic wafer is shown in Fig.~\ref{fig:f2_misalignment}b, along with the fitted normal distribution, from which we find a misalignment $\Delta~=~(9\pm 46$) nm.~\footnote{For comparison, using reflected light to image the alignment markers resulted in a misalignment of ($-3\pm 169$)~nm}  We attribute the slight mean misalignment of 9 nm to a rigid shift introduced during the nanofabrication, which is within the 15 nm layer alignment accuracy of the electron beam lithography system.  The spread of the distribution (46 nm standard deviation) mainly arises from imperfections within our imaging system, for example a beam-offset introduced by the bandpass filter that allows us to measure emission from the QDs.  To "successfully" couple the QD to a nanophotonic structure we require the total error to be smaller than the size of the features in its light-field.  Our total error, which is dominated by the random error of the procedure, is sufficiently small to enable excellent coupling to a PhCW.~\cite{Javadi2017}

A similar analysis of the nanoguides on the doped sample reveals the benefit of taking all images using the same filter.  The misalignment histogram for this sample, which we show in Fig.~\ref{fig:f2_misalignment}c, reveals a $\Delta$ of only ($1\pm 33$) nm.  That is, in this case we observe a vanishing average systematic shift of the alignment, and a smaller spread in the misalignment relative to that of the undoped sample.  We attribute this improvement to using the same physical imaging optics for both alignment and QD frames. Finally, since the beam positioning resolution of our state-of-the-art electron beam lithography is only 0.1 nm, we conclude that it is the optical aspect of our technique, and not the nanofabrication, which determines the final alignment precision.

\section{Effects of nanofabrication}\label{sec:Effects}
The effects of nanofabrication on the intrinsic optical properties of quantum dots are largely unknown. In fact, only the linewidth changes~\cite{Liu2018} and spectral shifts~\cite{Kaganskiy2015} of quantum dots in undoped micropillars have been studied; no such reports exist at all for high-quality, electronically contacted QDs.  To address this need, we record the fluorescence spectra from the QDs before and after nanofabrication, for both types of wafers by optically exciting them from the top.   For the QDs in the doped samples, we maintain the same bias voltage of 300~mV before and after nanofabrication.

\begin{figure}
\begin{center}
\includegraphics[width=8.4cm]{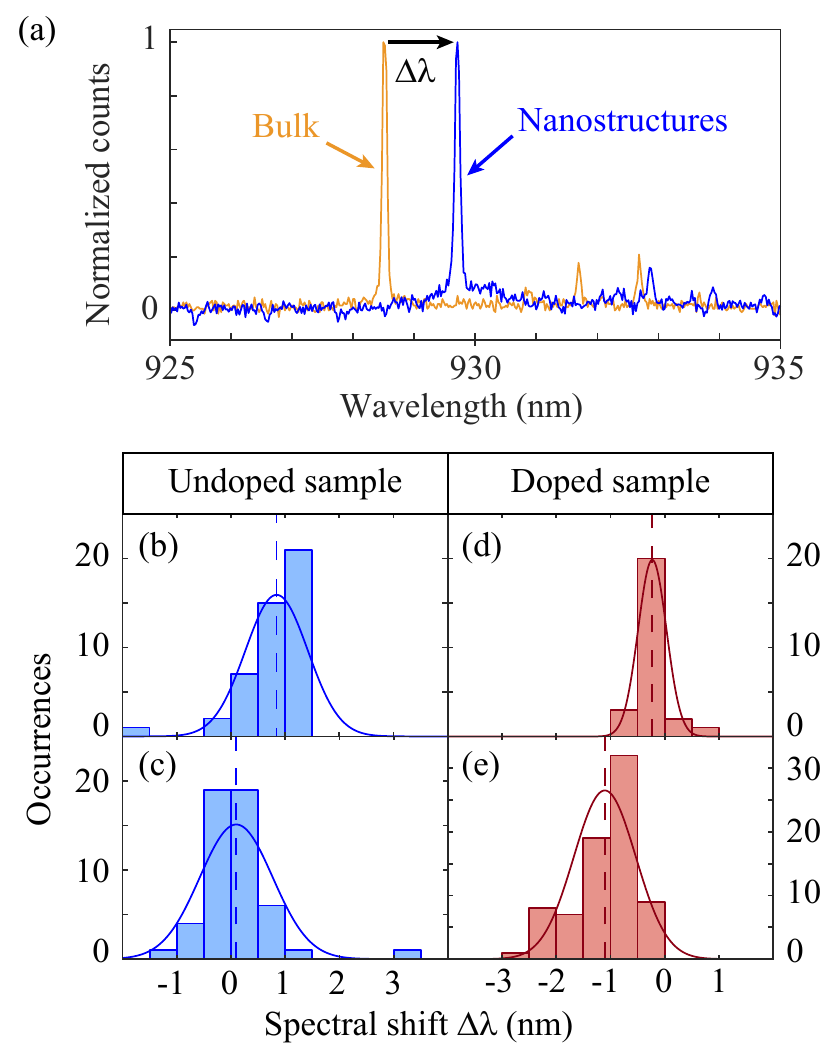}
\caption{Effects of nanofabrication on the spectral properties of QDs.  (a) Emission spectra for a QD both before nanofabrication (bulk sample), and after when it is embedded in a nanoguide.  Each spectra is normalized to its respective maximum, and the spectral shift $\Delta \lambda$ is marked.  (b) and (c) Histograms of spectral shifts for QDs embedded in nanobeam and PhCWs, respectively, in undoped samples.  Also shown are the fitted normal distributions from which we extract the mean $\Delta \lambda$ for each set of structures. (d) and (e) Same as in (b) and (c) but for the doped wafer, with all spectra taken at an applied $V = 300$~mV.}
\label{fig:f3_shifts}
\end{center}
\end{figure}

Exemplary spectra of the same QD in bulk and in a nanoguide, here in an undoped wafer, are shown in Fig.~\ref{fig:f3_shifts}a.
For this QD, we observe a clear shift $\Delta \lambda\approx1$~nm, which is a typical value for this wafer and this structure.  In fact, from the histogram of such shifts (Fig.~\ref{fig:f3_shifts}b) we calculate a mean $\Delta \lambda = \left(0.8\pm 0.6\right)$~nm.  Similar data for the QDs in PhCWs reveal a smaller $\Delta \lambda = \left(0.1\pm 0.7\right)$~nm (Fig.~\ref{fig:f3_shifts}c).  Shifts of this order of magnitude are consistent with either the creation of surface states during the nanofabrication,~\cite{Lee2006a} or changes to the stress and strain within the GaAs membrane due to the removal of the sacrificial layer.~\cite{Nakaoka2003} As we discuss below, the uniformity of the shifts suggests that the latter effect dominates, in which case the different values of $\Delta \lambda$ for the two types of structures may arise from their different dimensionalities and material compositions.

We perform similar experiments and analysis on the QDs embedded in nanophotonic waveguides on the doped samples. In this case, we measure $\Delta \lambda = \left(-0.2\pm 0.3\right)$~nm and $\left(-1.1\pm 0.6\right)$~nm for the QDs embedded in the nanoguides and PhCWs, respectively (Figs.~\ref{fig:f3_shifts}d and e). Although these shifts are of the same magnitude as those in the intrinsic sample, they are now in the opposite direction, demonstrating that the fabrication process affects the layered and homogeneous wafers in a different manner.

\begin{figure}
\begin{center}
\includegraphics[width=8.4cm]{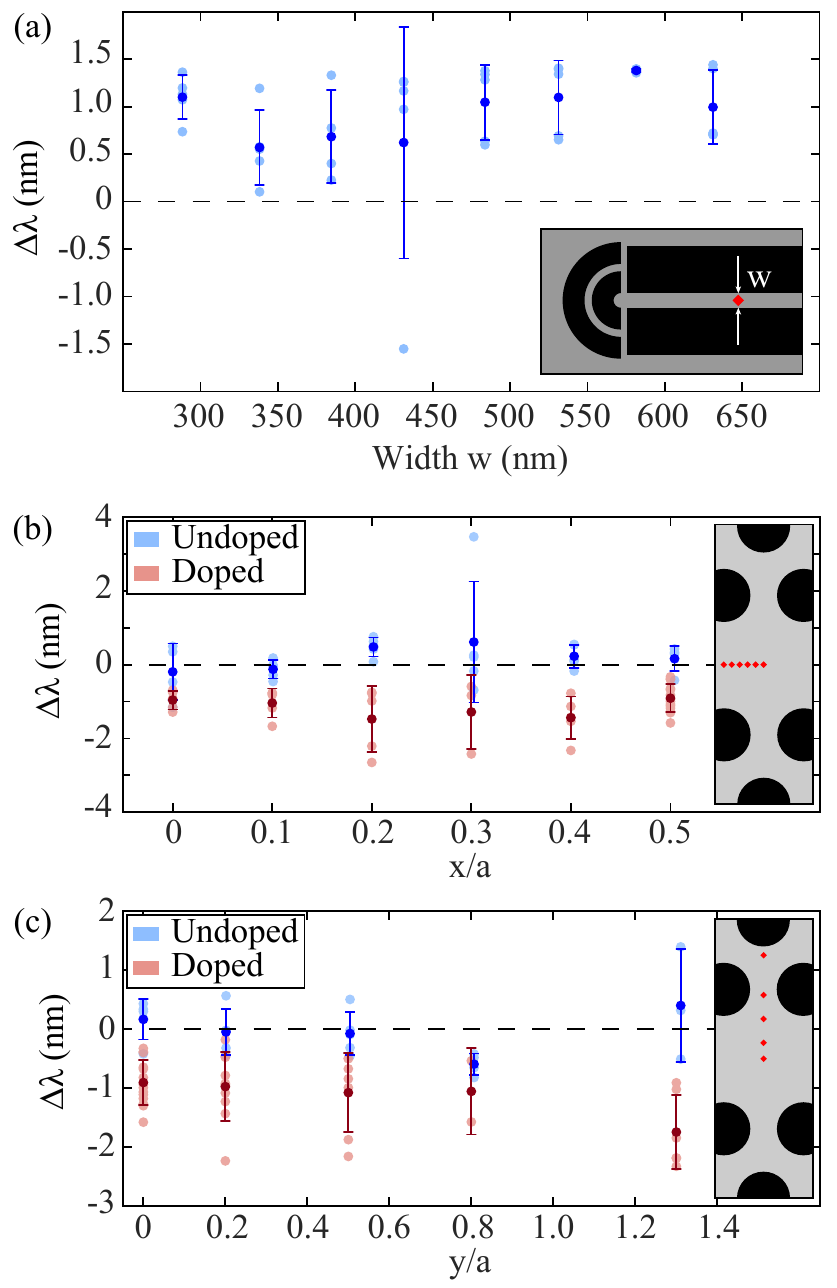}
\caption{Dependence of the spectral shift of QD resonances due to integration in (a) nanoguides of various sizes and (b) and (c) different positions within PhCWs.  The dark colored symbols and error bars are the mean and standard deviation of $\Delta\lambda$, with the corresponding individual measurements shown by the faded circles. The inset to (a) shows the width of the nanoguide, while in (b) and (c) the positions of the QDs within the unit cell of the PhCW are depicted.}
\label{fig:f4_ShiftsVDistance}
\end{center}
\end{figure}

The spectral shifts presented in Fig.~\ref{fig:f3_shifts} can be further subdivided according to nanoguide width or emitter position within the unit cell for the PhCWs. The results, presented in Fig.~\ref{fig:f4_ShiftsVDistance}, demonstrate that in either case the shift is constant to within the measurement error.
This means that, for the nanoguides (Fig.~\ref{fig:f4_ShiftsVDistance}a), there is no appreciable difference to $\Delta\lambda$ between QDs that are 144~nm away from the waveguide wall and those with a separation of 315.5~nm.  This is likewise true for the QDs in the PhCWs, regardless of whether the QD was shifted in $x$ or $y$ along the unit cell (Figs.~\ref{fig:f4_ShiftsVDistance}b and c, respectively), although a larger $\Delta\lambda$ is observed for the doped sample.  Here, the separation between the QDs and the nearest edge of the PhCWs varied between about 29 and 171 nm.  The uniformity of these shifts allows us to conclude that they do not arise due to the presence of surfaces, for example due to the trapping of charges, but rather supports the notion that their origin can be traced to the general relaxation of the GaAs membrane due to the removal of the sacrificial AlGaAs layer.

\begin{figure}
\begin{center}
\includegraphics[width=7.8cm]{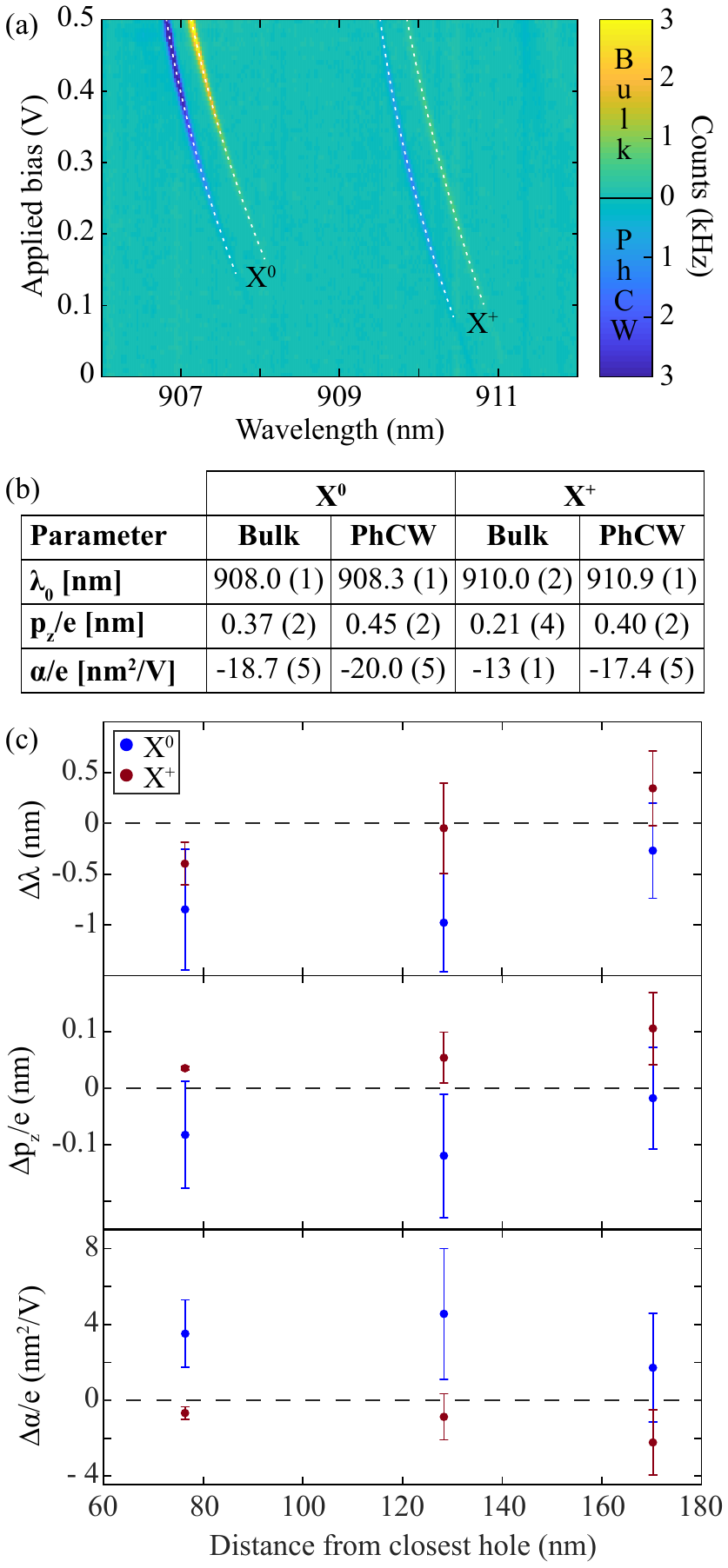}
\caption{Electrical tuning of QDs before and after fabrication of a PhCW. (a) Voltage-wavelength PL spectra for a QD located 170 nm away from the nearest surface of a PhCW. The fluorescence plateaus of both the neutral $\left(X^0\right)$ and charged $\left(X^+\right)$ excitons are seen in the maps and, in this exemplary case, both blue shift by about 0.3 nm with applied bias. The curvature of the plateaus, however, changes differently, revealing that the nanofabrication affects the two excitons differently. These changes can be quantified via a theoretical model (dashed line), as discussed in the main text. (b) Fit parameters for the different excitons shown in (a).  (c) Nanofabrication induced changes to the Stark-shift parameters as a function of nominal distance to the nearest vertical surface.  Two to four QDs were measured at each position of the PhCW unit cell, and the error bars represent the calculated standard deviation.}
\label{fig:f5_VFmap}
\end{center}
\end{figure}

Encouragingly, the fabrication-induced shifts can be largely overcome through the electrical gating of the QDs in the doped sample.  This can be seen in the $\mu$PL spectra, taken with above-band excitation for different applied bias voltages, an example of which is shown in Fig.~\ref{fig:f5_VFmap}.
Here the charge plateaus for both the neutral $\left(X^0\right)$ and charged $\left(X^+\right)$ excitons are displayed both in bulk and after integration into a PhCW. The charged exciton is identified as positively charged because it appears at lower applied bias voltages (i.e. larger fields) than the neutral excitonic line.~\cite{Kirsanske2017, Dalgarno2008} For both excitons, we observe a blue-shift of the emission wavelength due to the nanofabrication, whose statistics were captured in Fig.~\ref{fig:f3_shifts}e.  It is clear that, for this QD, increasing the applied bias by about 150 mV recovers the bulk emission wavelength over a fairly broad bandwidth of $\approx 0.6$ and $0.5$~nm for $X^0$ and $X^+$, respectively.

By fitting the Stark shift of the QD as is done in Fig.~\ref{fig:f5_VFmap}a (dashed lines), we quantify the effects of the nanofabrication on the emitter dipole. This model describes the quadratic dependence of the Stark shift on the transition energy,~\cite{Fry2000, Finley2004, Bennett2010} written in the terms of emission wavelength as
\begin{equation}\label{eq:SS}
\frac{hc}{\lambda} = \frac{hc}{\lambda_0} - p_z F + \alpha F^2.
\end{equation}
Here, $\lambda_0$ is the QD emission wavelength at vanishing applied field, $p_z$ is the permanent dipole moment of the QD in the growth direction, $\alpha$ is the polarizability of the emitter, and $F = \left(V-V_i\right)/t$ is the applied field for a given bias voltage $V$. The thickness of the intrinsic layer surrounding the QDs is nominally $t=70$~nm and the built-in field has been calculated from the difference between the Fermi levels of the p- and n-doped GaAs layers,~\cite{Sze2006} resulting $V_i/t = 224.24$~kV/cm. The fit parameters for the 4 excitonic lines shown in Fig.~\ref{fig:f5_VFmap}a are given in Fig.~\ref{fig:f5_VFmap}b, and are comparable with values reported in literature.~\cite{Bennett2010, Finley2004, Warburton2002, Hsu2001}

We analyze similar frequency-voltage spectral maps for QDs located at different positions within the PhCW unit cell, quantifying possible fabrication induced changes to the dipole parameters. The results are presented in Fig.~\ref{fig:f5_VFmap}c. This analysis reveals that the effects of the nanofabrication on the charged exciton are generally small compared to the changes of $X^0$.  This observation may be related to the different exciton wavefunctions associated with neutral and charged excitons where the former is further extended than the latter,~\cite{Finley2004} therefore possibly making it more susceptible to the local environment. We also note that no pronounced dependence on the distance of the QDs from the etched holes are observed, which indicates once again that strain and stress alterations rather than surface defects may be responsible for the observed spectral changes. 

\section{Conclusions and outlook}
We have presented a method for precisely locating epitaxially grown QDs that allows us to deterministically integrate the emitters into nanostructured photonic waveguides.  In contrast to previous approaches, we only rely on photoluminescence data and not on reflections to image both the QDs and a global reference frame.  This improvement enables us to position high-quality, gate-tunable QDs with a random error of only 33~nm, almost halving the error of previously reported $\mu$PL results.~\cite{Thon2009}  Here, we employ this method to couple QDs to the hot-spot of a PhCW with a yield of over 90\%, and can even place electronically-contacted emitters within about 30 nm from a hole with a 33\% success rate.  Our method allows us to investigate the effects of nanofabrication on the emission wavelength and exciton properties of individual QDs, and will enable similar studies of other emission properties such as QD linewidths and coherence.

The obtained precision suffices next-generation quantum nanophotonics experiments with waveguides and cavities, such as mapping out the spatial dependence of the optical local density of states or precisely probing the spatial polarization profile of nanophotonic waveguides leading to chiral emission,~\cite{Lodahl2017} or robust demonstration of strong coupling.~\cite{Kuruma2016} In fact, in a subsequent work,~\cite{Chu2019} we make use of the technique presented here to spatially and spectrally position QDs in PhCWs. In doing so, we are able to simultaneously exploit both the slow-light effect and the high confinement of the propagating mode to overcome intrinsic non-radiative processes and significantly boost the device quantum efficiency. Combined with the demonstrated spectral control via Stark tuning, a path is laid out towards the deterministic coupling of multiple QDs via the engineered dipole-dipole interaction through the waveguide. Such controlled interaction may be applied for two-qubit gates between emitting QDs, enabling the generation of advanced photonic quantum resources such as 2D clusters of multiple entangled photons.~\cite{Buterakos2017}

\begin{acknowledgments}
The authors gratefully acknowledge Sandra {\O}. Madsen for her help in the construction of the optical setup and Leonardo Midolo and Zhe Liu for assistance in fabrication. We gratefully acknowledge financial support from the Danish National Research Foundation (Center of Excellence Hy-Q), the Europe Research Council (ERC Advanced Grant SCALE), the European Union's Horizon 2020 research and innovation program under the Marie Sk\l{}odowska Curie grant agreement no. 753067 (OPHOCS), Innovation Fund Denmark (Quantum Innovation Center Qubiz), and the Danish Research Infrastructure Grant (QUANTECH). R.S., A.L. and A.D.W. acknowledge gratefully support of DFG-TRR160, BMBF - Q.Link.X 16KIS0867, and the DFH/UFA CDFA-05-06.
\end{acknowledgments}

\section*{data availability}
The data that support the findings of this study are available from the corresponding author upon reasonable request.

\appendix
\section{Optical setup}{\label{sec:Setup}}
\begin{figure}
\begin{center}
\includegraphics[width=8.5cm]{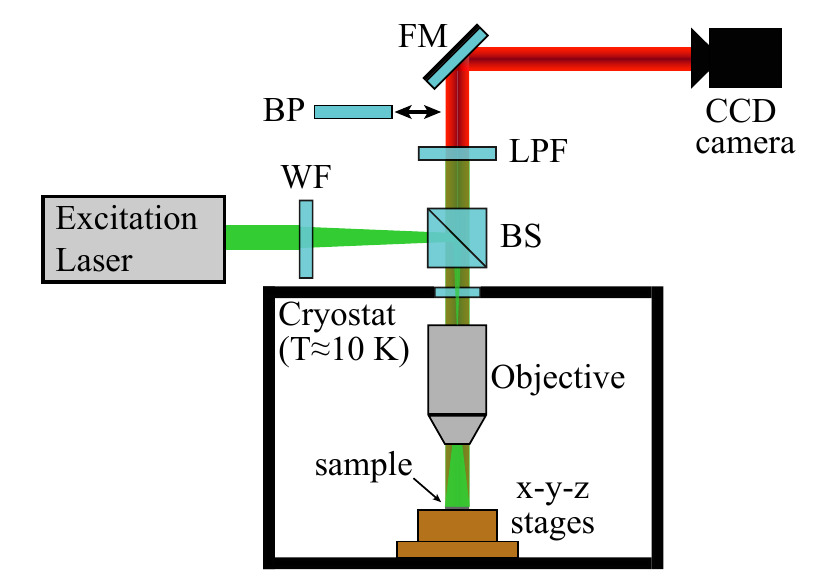}
\caption{Schematics of the optical setup used for the image acquisition procedure (WF~=~wide-field lens, BS~=~beam-splitter, LPF~=~long-pass filter, BP~=~band-pass filter, FM~=~flipping mirror).
\label{fig:f6_Setup}}
\end{center}
\end{figure}
The optical setup employed to acquire the images for locating both reference markers and QDs is sketched in Fig.~\ref{fig:f6_Setup}. The sample is mounted in a closed-cycle cryostat containing three piezo-stages and a microscope objective (magnification~=~100x and NA~=~0.85) inside the vacuum chamber. The excitation laser ($\lambda$~=~780 nm) is coupled to the objective through a 10:90 (reflection:transmission) beam-splitter (BS). A wide-field (WF) lens focuses the beam onto the back focal plane of the objective in order to achieve a large illumination area and cover the whole field of view. The light emitted from the sample is sent through a long-pass filter (LPF) and collected by a CCD camera (Andor iKon-M), with a $13.3\times13.3~\text{mm}^2$ sensor formed by $1024\times1024$ active pixels. The final field-of-view covers an area of about $60\times60~\text{$\mu$m}^2$, which means that every pixel corresponds to a region of about $59\times59~\text{nm}^2$ of the sample. In  the case of the intrinsic sample, a band-pass (BP) filter is introduced in the collection path to switch between the two different types of acquired images. Furthermore, we adjust the acquisition time to avoid any relevant drift of the system while achieving a high signal-to-noise ratio: typically, 1~s for the reference markers and 1 to 10 s for the QD images. Finally, the optical setup can be easily converted into a confocal configuration by moving the wide-field lens out of the excitation path and by redirecting the emission signal towards a spectrometer.

\section{Localization of the reference markers}{\label{sec:Loc_Xs}}
\begin{figure}
\begin{center}
\includegraphics[width=8.5cm]{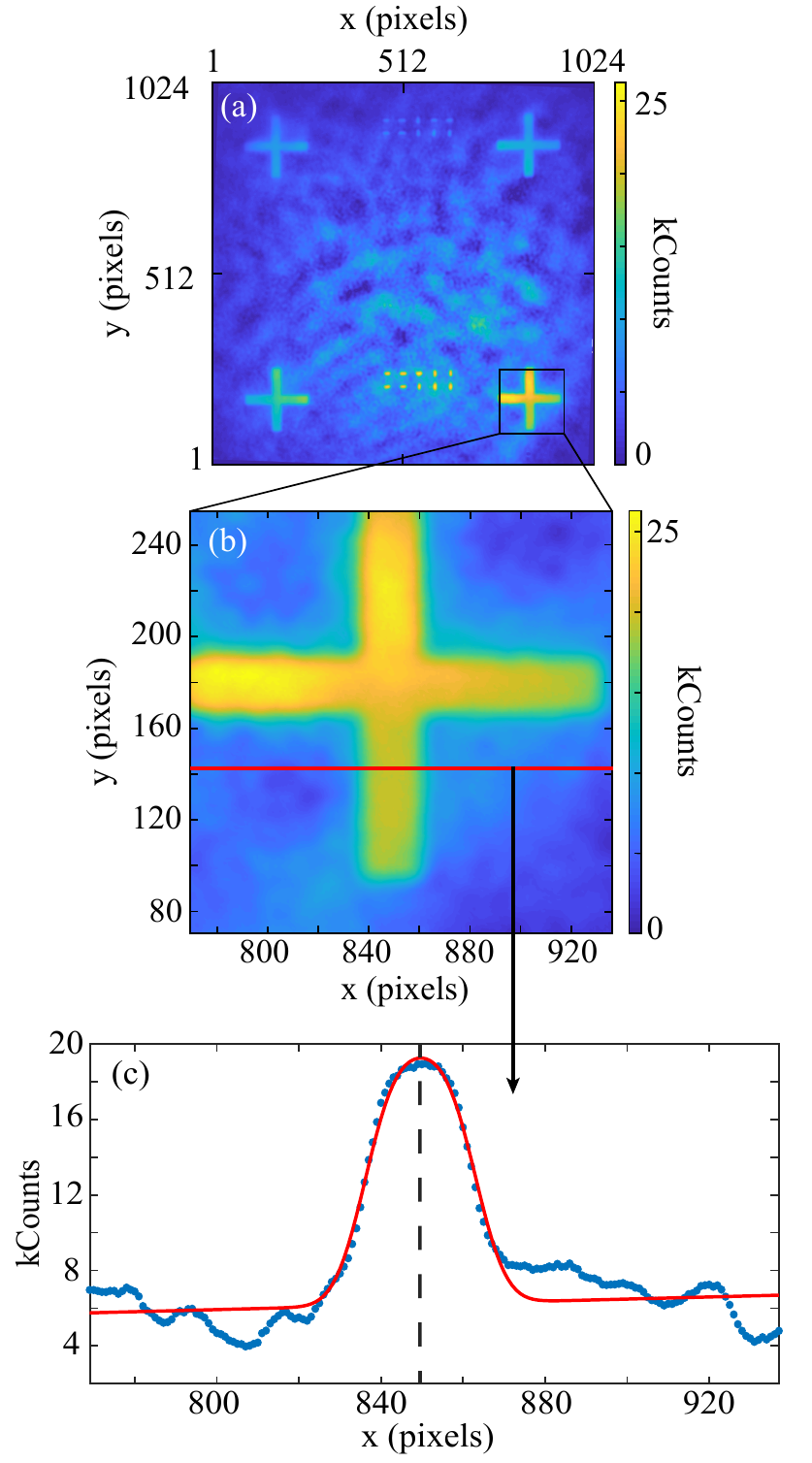}
\caption{Summary of the process for the localization of the reference markers. (a) Original image, after rotation and background correction. (b) Cropped area around the reference marker marked in (a). (c) Plot of the intensity detected by the pixels marked by the red line in (b). The blue points show the data, while the red line is the fitting curve used to calculate the coordinate of the center x$_{c,i}$ (indicated by the black dashed line).}
\label{fig:f7_fitMarkers}
\end{center}
\end{figure}
The procedure to find the spatial position of the reference markers is summarized in Fig.~\ref{fig:f7_fitMarkers}. Every acquired image initially undergoes two correction steps in order to facilitate the subsequent analysis: (i) a rotation, to align the crosses to the frame of the picture, and (ii) a background subtraction, to remove the Gaussian-shaped intensity distribution generated by the photoluminescence of the substrate (cf. panel II of Fig.~\ref{fig:f1_concept}). An example of the resulting images is shown in Fig.~\ref{fig:f7_fitMarkers}a. The areas around each alignment crosses are subsequently cropped from the original image and analyzed independently (Fig.~\ref{fig:f7_fitMarkers}b).

The center of each reference cross is identified as the intersection between the central axes of the horizontal and vertical arms, which are calculated as linear fits of the central positions of multiple cross-sections analyzed for each arm. Fig.~\ref{fig:f7_fitMarkers}c demonstrates an example of such cross-sections, as it plots the photoluminescence intensity detected along the pixels marked by the red line in Fig.~\ref{fig:f7_fitMarkers}b. The signal is fitted with the function
\begin{widetext}
\begin{equation}
y\left(x\right) = A\cdot\frac{erf{\left(\sqrt{\frac{1}{2\,\sigma}}\,\left(\text{x}_{c,i}-\frac{d}{2}-x\right)\right)} - erf{\left(\sqrt{\frac{1}{2\,\sigma}}\,\left(\text{x}_{c,i}+\frac{d}{2}-x\right)\right)}}{erf{\left(\sqrt{\frac{1}{2\,\sigma}}\,\left(-\frac{d}{2}\right)\right)} - erf{\left(\sqrt{\frac{1}{2\,\sigma}}\,\left(\frac{d}{2}\right)\right)}}\,+B\cdot{x}\,+C,
\label{eq_XsFit}
\end{equation}
\end{widetext}
which is the result of the convolution between a 1D Gaussian $g\left(x\right) = \exp{\left(-x^2/2\sigma\right)}$ and the rectangular function $\Pi\left(x\right)$
\begin{equation}
\Pi\left(x\right) = \begin{cases} 1, & -\frac{d}{2} \leq x\leq \frac{d}{2} \\ 0, & \left|x\right|>\frac{d}{2}, \end{cases}
\end{equation}
where $d$ is the nominal value of the arm width. Moreover, $erf{\left(x\right)} = \frac{2}{\sqrt{\pi}}\int_{0}^{x} e^{-t^2} dt$ is the error function, the parameter $A$ is the amplitude of the fitting function, and the linear term $B\cdot x$ and the constant $C$ account for any spatial-dependent background contribution that has not been completely corrected for by the initial steps. The central position of the arm in the $i$th cross-section $x_{c,i}$ is evaluated as one of the fit parameters in eq.~\eqref{eq_XsFit} and its uncertainty is calculated as half of the 95.4\% confidence interval (i.e. 2 standard deviations) obtained from the fit. This localization procedure is repeated for every isolated cross in order to find the coordinates of their centers.

\section{Localization of QDs}{\label{sec:Loc_QDs}}
\begin{figure}
\begin{center}
\includegraphics[width=8.5cm]{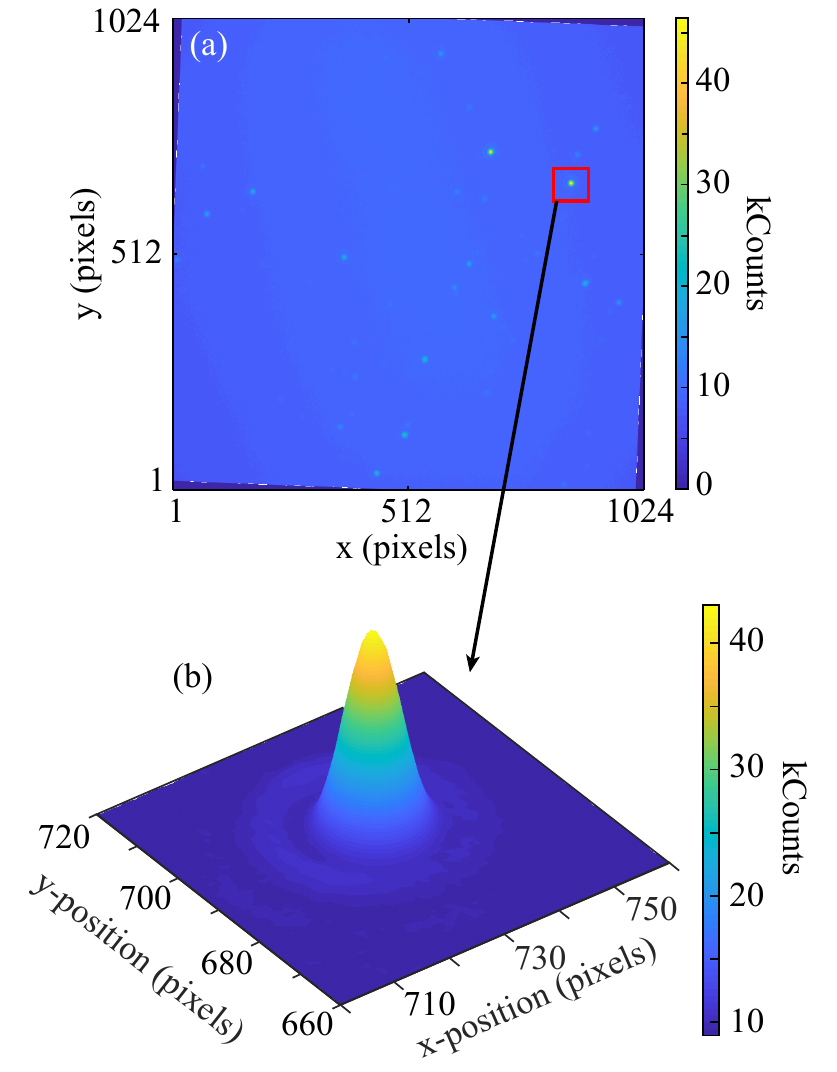}
\caption{Localization of the QDs. (a) Original image, after rotation correction. The QDs appear as tiny bright spots on a uniform dark background. (b) Surface plot of the intensity spatial distribution in one the cropped regions of interest.}
\label{fig:f8_fitQDs}
\end{center}
\end{figure}
Fig.~\ref{fig:f8_fitQDs} outlines the procedure to locate the QD centers. After the acquired image is corrected by applying the same rotation used for the related picture with alignment markers, small regions of interest are defined around the bright spots that identify the QDs (Fig.~\ref{fig:f8_fitQDs}a). The typical intensity distribution inside each one of the selected regions is shown in Fig.~\ref{fig:f8_fitQDs}b and demonstrates the expected Airy pattern. Under the conditions used to take the images in this work, the emission pattern is well-fitted by a 2D Gaussian and the coordinates of the detected maximum correspond to the location of the point source.~\cite{Mortensen2010}

As the point spread function that describes the intensity distribution is known, super-resolution techniques can be employed to determine the uncertainty on the position of the QD. In fact, the variance of each of the detected coordinates describing the position of a single point-source can be written as~\cite{Mortensen2010}
\begin{equation}
\sigma_{tot, x}^2 = \frac{\sigma_{a,x}^2}{N}\left(\frac{16}{9} + \frac{8\pi\sigma_{a,x}^2b^2}{Na^2}\right),
\label{eq_QDunc}
\end{equation}
where the subscript $x$ indicates that it is calculated for the x-coordinate, but the same formula can be written also for the y-coordinate. In eq.~\eqref{eq_QDunc}, $\sigma_{a,x}^2 = \sigma_x^2 + a^2/12$, with $a^2$ is the pixel area and $\sigma_x$ is the value of the standard deviation calculated for the fitting 2D Gaussian along the x-direction. Since the investigated intensity distribution is written in terms of pixels, we set $a^2=1$. The parameter $N$ describes the total number of photons that are emitted by the QD and corresponds to the volume under the 2D Gaussian. The variable $b^2$ indicates the background level of the analyzed region. As all the parameters in eq.~\eqref{eq_QDunc} are known, the uncertainty on the detected QD location can be calculated as $\delta_x = \sqrt{\sigma_{tot, x}^2}$.

\section{Evaluation of the final alignment accuracy}{\label{sec:Loc_Wgs}}
\begin{figure}
\begin{center}
\includegraphics[width=8.5cm]{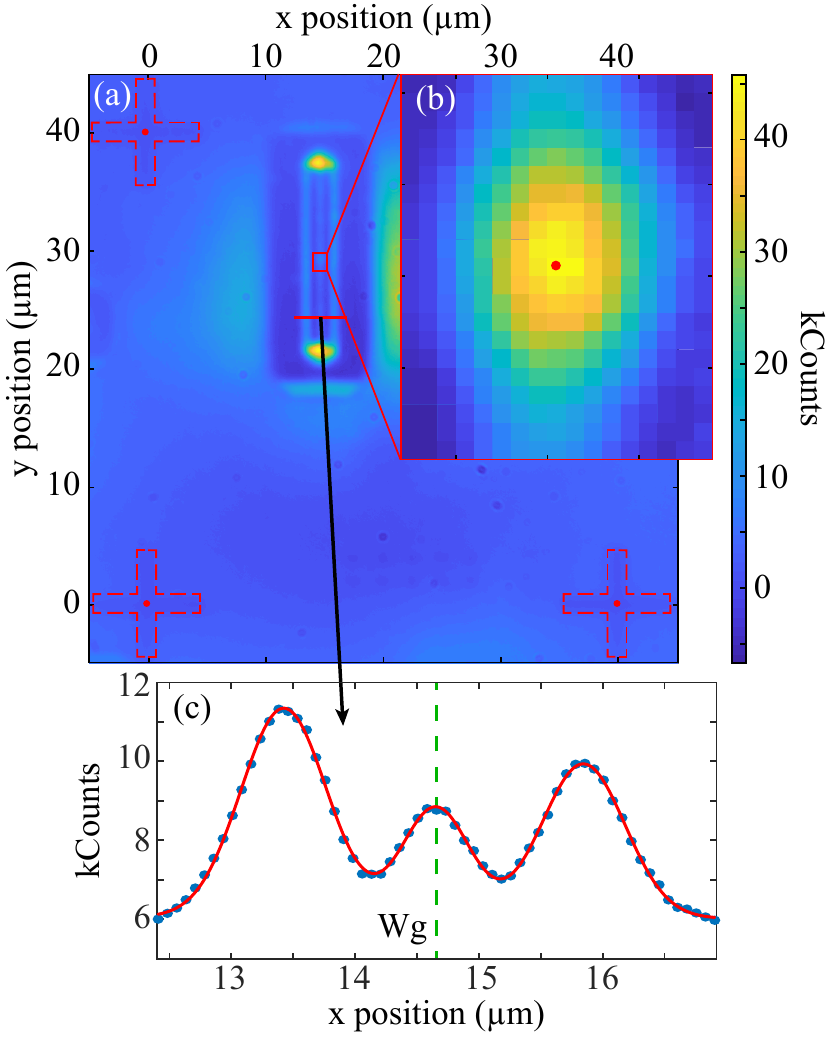}
\caption{Characterization of the final alignment accuracy. (a) Example of the intensity map analyzed for locating the fabricated nanoguides, obtained by using the wetting layer emission as illumination source. The positions of the reference markers (outlined as a cross defined by a red dashed lines) are detected by the same localization procedure as used before. (b) Selected region of interest about the QD embedded in the nanoguide, which is cropped from the acquired image where only the QD emission is visible.  The center of the emitter (marked as a red dot) is found by fitting the emission pattern with an elliptical Airy function.  (c) Typical cross-section of a nanoguide, taken along the red line displayed in (a). In the plot, the blue dots are the counts detected at each pixel of the CCD camera, whereas the red curve is the result of the fit. The green dashed line marks the center of the waveguide Wg.}
\label{fig:f9_fitWg}
\end{center}
\end{figure}
In order to quantitatively characterize the misalignment $\Delta$ between the fabricated nanoguides and the pre-selected QDs, we follow the same approach used for the initial localization of the quantum emitters and thus acquire two sets of images: one for the localization of reference markers and nanostructures, where the emission from the wetting layer is used as the source of illumination, and another for the emitters. Fig.~\ref{fig:f9_fitWg} outlines the main steps of the procedure. After employing the fitting routine explained in Appendix~\ref{sec:Loc_Xs} to detect the position of the reference markers (shown as red dashed crosses in Fig.~\ref{fig:f9_fitWg}a), we determine the position of the central axis of the nanoguides by fitting multiple cross-sections of the nanostructures, which appear blurred in the acquired images since their width is smaller than the diffraction limit of our optical setup. Nevertheless, their profile is still visible and can be easily identified as the central Gaussian peak in each cross-section, as demonstrated in Fig.~\ref{fig:f9_fitWg}c. The two side peaks are generated by the light that is scattered from the external edges of the trenches on both sides of the suspended nanoguide (cf. the inset in Fig.~\ref{fig:f2_misalignment}a). Fitting the intensity distribution across the waveguide allows us to extract the central position of every cross-section (marked as a green dashed line in Fig.~\ref{fig:f9_fitWg}c), with an uncertainty calculated as half of the 95.4\% confidence interval and resulting on average of about 25~nm. Finally, the center of the investigated nanostructure is evaluated as a weighted average of each calculated position. We note that we could determine only y-coordinates (x-coordinates) from nanoguides aligned along the x-direction (y-direction) due to the orientation of the corresponding fitted cross-sections.

The positions of the QDs are found with a procedure similar to the one described in Appendix~\ref{sec:Loc_QDs}. The presence of suspended nanostructures, however, distorts the emission pattern of the embedded emitters, 
which now resembles a 2D Gaussian with a cross-section that is elliptical rather than circular, as shown in Fig.~\ref{fig:f9_fitWg}b. Moreover, the light generated by the QDs propagates also through the nanoguide and is scattered towards the camera from the out-couplers, the waveguide itself and the surrounding substrate. The acquired images thus present a reduced signal-to-background ratio, which increases the uncertainty of the QD position. In this situation, the considerations used in Appendix~\ref{sec:Loc_QDs} do not hold any more and we therefore decided to use the more accurate Airy function to best fit the data. The uncertainty on the QD position is now evaluated as half of the 95.4\% confidence interval obtained from the fit and results comparable to the value obtained for the nanoguides. The final misalignment $\Delta$ between the central axis of the nanostructures and the center of the embedded QDs (marked as a red dot in Fig.~\ref{fig:f9_fitWg}b) is calculated by subtracting the values of analogous coordinates.

\printfigures

\end{document}